\documentclass[conference]{IEEEtran}
\usepackage{fancyhdr}
\ifCLASSINFOpdf
   \usepackage[pdftex]{graphicx}
\else
\fi
\usepackage[cmex10]{amsmath}
\usepackage{amssymb}
\usepackage[caption=false,font=footnotesize]{subfig}
\usepackage{url}

\usepackage{fancyvrb}
\VerbatimFootnotes

\usepackage{flushend}

\usepackage[utf8]{inputenc}
\usepackage{algorithmic}

\usepackage{xspace}
\newcommand{\lqcd}{Lattice QCD\xspace}

\newcommand{\KNC}{Intel\textsuperscript{\textregistered} Xeon Phi\textsuperscript{\texttrademark}~co-processor\xspace}

\newcommand{\knc}{KNC\xspace}
\newcommand{\kncs}{KNCs\xspace}
\newcommand{\knl}{KNL\xspace}

\providecommand{\norm}[1]{\lVert#1\rVert}
\DeclareMathOperator{\ceil}{ceil}

\begin{document}

\title{Lattice QCD with Domain Decomposition on Intel\textsuperscript{\textregistered} Xeon Phi\textsuperscript{\texttrademark}~Co-Processors}




%
\author{\IEEEauthorblockN{
Simon Heybrock\IEEEauthorrefmark{1},
Bálint Joó\IEEEauthorrefmark{2},
Dhiraj D. Kalamkar\IEEEauthorrefmark{3},
Mikhail Smelyanskiy\IEEEauthorrefmark{4},\\
Karthikeyan Vaidyanathan\IEEEauthorrefmark{3},
Tilo Wettig\IEEEauthorrefmark{1},
and
Pradeep Dubey\IEEEauthorrefmark{4}
}
\IEEEauthorblockA{\IEEEauthorrefmark{1}Institute for Theoretical Physics, University of Regensburg, Germany}
\IEEEauthorblockA{\IEEEauthorrefmark{2}Thomas Jefferson National Accelerator Facility, Newport News, VA, USA}
\IEEEauthorblockA{\IEEEauthorrefmark{3}Parallel Computing Lab, Intel Corporation, Bangalore, India}
\IEEEauthorblockA{\IEEEauthorrefmark{4}Parallel Computing Lab, Intel Corporation, Santa Clara, CA, USA}
}


\maketitle
\thispagestyle{fancy}
\lhead{}
\rhead{}
\chead{}
\lfoot{\footnotesize{
    \vspace{-2mm}
SC14, November 16-21, 2014, New Orleans, Louisiana, USA,
\newline accepted for publication, \copyright 2014 IEEE, DOI 10.1109/SC.2014.11}} 
\rfoot{} 
\cfoot{} 
\renewcommand{\headrulewidth}{0pt}
\renewcommand{\footrulewidth}{0pt}

\begin{abstract}


      The gap between the cost of moving data and the cost of computing
  continues to grow, making it ever harder to design iterative solvers on
  extreme-scale architectures.  This problem can be alleviated by
  alternative algorithms that reduce the amount of data movement.  We
  investigate this in the context of Lattice Quantum Chromodynamics
  and implement such an alternative solver algorithm, based on domain
  decomposition, on \KNC (\knc) clusters.  We demonstrate
  close-to-linear on-chip scaling to all 60 cores of the \knc.  With a mix of
  single- and half-precision the domain-decomposition method sustains 400-500~Gflop/s
  per chip.  Compared to an optimized \knc implementation of a
  standard solver~\cite{Joo:2013a}, our full multi-node domain-decomposition solver
  strong-scales to more nodes and reduces the time-to-solution by a factor of 5.

\end{abstract}




%
\IEEEpeerreviewmaketitle

\section{Introduction}


It is well known that the cost of moving data exceeds by far the cost of computing (see, e.g., Sec.~5.2 of Ref.~\cite{DARPA:2008}), and this gap is constantly growing.
This poses serious challenges for the design of iterative solvers on extreme-scale architectures.
For example, the most time-consuming part of Lattice Quantum Chromodynamics (\lqcd) is the iterative solver of the discretized Dirac equation.
The \lqcd community has always shown strong interest in porting their code to new and more powerful architectures.
Many recently installed clusters couple commodity processors to co-processors such as the Intel\textsuperscript{\textregistered} Xeon Phi\textsuperscript{\texttrademark} `Knights Corner' co-processor (\knc).\footnote{Intel, Xeon, and Intel Xeon Phi are trademarks of Intel Corporation in the U.S. and/or other countries.}
A first implementation of a \lqcd solver for the \knc was introduced in Refs.~\cite{Joo:2013a} and \cite{Vaidyanathan:2014a}.
The code presented there demonstrated that it is possible to obtain a high flop count on the \knc for a memory-bandwidth-bound stencil operator.
This was facilitated by a 3.5D blocking scheme~\cite{Nguyen:2010} and an optimized prefetch pattern based on a custom-designed code generator.
Furthermore, good multi-node performance was achieved, as long as the local volume per \knc was large enough to offset the overhead of boundary communications and global reductions.

Despite this success it is desirable (and in some cases even necessary) to surpass the bounds given by the memory bandwidth and to improve the strong-scaling behavior.
In our assessment, a considerable improvement is only possible with a different algorithm that moves less data both from and to main memory and between different \kncs than a standard solver does.
We therefore work with a solver based on domain decomposition (DD) which has exactly these properties.
In this paper we describe the implementation of a DD method for \lqcd on \knc-based architectures and the aspects relevant for achieving very high performance both on a single node and on multiple nodes.

This paper is organized as follows.
Section \ref{sec:algorithmic_considerations} serves as an extended introduction.
In Sec.~\ref{sec:knc} we outline the \knc architecture and in Sec.~\ref{sec:lqcd} we provide some relevant details about \lqcd.
We introduce iterative solvers in Sec.~\ref{sec:solvers} and motivate the use of a DD preconditioner, which is described in Sec.~\ref{sec:dd}.
We provide details of our implementation in Sec.~\ref{sec:implementation} and discuss the main aspects relevant for the \knc architecture.
Results are presented in Sec.~\ref{sec:results}, specifically those for the single-node implementation in Sec.~\ref{sec:results_single} and those for multi-node in Sec.~\ref{sec:results_multi}.
We discuss related work in Sec.~\ref{sec:related} and conclude in Sec.~\ref{sec:conclusion}.



\section{Algorithmic considerations}
\label{sec:algorithmic_considerations}
\IEEEpubidadjcol

\subsection{\knc architecture}
\label{sec:knc}

The \knc architecture directly influences algorithm choices.
It features $60$ or $61$ in-order cores on a single die.
Each core has 4-way hyper-threading support to help hide memory and multi-cycle instruction latency.
Instructions from a single thread can be issued only every other cycle, so a minimum of two threads per core is necessary to achieve full pipeline utilization.
To maximize area and power efficiency, these cores are less aggressive --- that is, they have lower single-threaded instruction throughput --- than CPU cores and run at a lower frequency.
However, their vector units are 512 bits wide and execute 8-wide double-precision or 16-wide single-precision single-instruction multiple-data (SIMD) instructions in a single clock.
The available \knc variants run at slightly above $1~\mathrm{GHz}$ and can then deliver up to around $1$ or $2~\mathrm{Tflop/s}$ floating-point performance in double- and single-precision, respectively.
The \knc has two levels of cache: a single-cycle 32~kB 1st level data cache (L1) and a larger globally coherent 2nd level cache (L2) that is partitioned among the cores.
Each core has a 512~kB partition.
The memory bandwidth in streaming can reach $150~\mathrm{GB/s}$ or slightly higher~\cite{Joo:2013a}.

The \knc has a rich instruction set architecture that supports many flavors of scalar and vector operations, including fused multiply-add.
To reduce the instruction foot-print, \knc operations support various in-flight transformations of one of the operands.
Most of the operations can take one of the operands from memory, and support a broadcast.
Load and store operations support up-conversion from 16-bit to 32-bit floating-point numbers and down-conversion from 32-bit to 16-bit.
In addition, the \knc has a dual-issue pipeline which allows prefetches and scalar instructions to co-issue with vector operations in the same cycle, to remove them from the critical path.


\subsection{\lqcd kernel}
\label{sec:lqcd}

\lqcd describes the interaction between quarks and gluons on an $N_d=4$ dimensional space-time lattice with lattice spacing $a$ and $V = L_x L_y L_z L_t$ sites, where $L_x$, $L_y$, $L_z$, and $L_t$ are the dimensions of the lattice in the $x$, $y$, $z$, and $t$ directions, respectively.
Quark fields (spinors) are ascribed to the sites of the lattice and carry $12$ complex degrees of freedom (3 color $\times$ 4 spin components, i.e., $24$ real components in total).
Gluon fields are ascribed to the links between sites and are represented by $\mathrm{SU}(3)$ matrices.
The interaction between quarks and gluons is given by the Dirac operator.
Here we use a formulation based on the Wilson discretization~\cite{Wilson:1975raey} of this operator, given by a matrix
\begin{align}
    A &= (N_d+m) -\tfrac{1}{2} D_{\text{w}} + D_{\text{cl}}
\end{align}
with a quark-mass parameter $m$ and the Wilson operator
\begin{align}
    D_{\text{w}} &= \sum_{\mu=1}^4 \left[ (1-\gamma_{\mu}) U_x^{\mu} \delta_{x+\hat{\mu},x'} + (1+\gamma_{\mu}) U_{x-\hat{\mu}}^{\mu\dagger} \delta_{x-\hat{\mu},x'} \right].
\end{align}
The sum is over the four space-time directions, $U_x^{\mu}$ is the gauge link matrix connecting site $x$ with its neighbor in direction $\mu$, and the $4\times4$ matrices $\gamma_{\mu}$ are elements of the Dirac spin-algebra.
The Wilson operator is a 9-point nearest-neighbor stencil in four dimensions, with 24 internal degrees of freedom.
Furthermore, we use the Clover correction term $D_{\text{cl}}$, given by
\begin{align}
    D_{\text{cl}} &= c_{\text{sw}} \sum_{\mu,\nu=1}^{4} \frac{i}{8} \sigma_{\mu\nu} \hat{F}_{\mu\nu},
\end{align}
which removes the discretization errors of order $a$ from the Wilson operator $D_{\text{w}}$ if the real parameter $c_{\text{sw}}$ is tuned appropriately.
$\sigma_{\mu\nu}$ is a combination of the $\gamma$-matrices.
The field strength tensor $\hat{F}_{\mu\nu}$ is a certain sum over products of gauge link matrices and resembles the shape of a four-leaved clover.
Hence the name of the term $D_{\text{cl}}$.
The remaining errors in the Wilson-Clover operator $A$ are of order $a^2$ or higher.
Details on the construction of the Clover term can be found in Ref.~\cite{Sheikholeslami:1985ij}.
For our purposes its relevant properties are that (1) it is a local term, coupling only the $24$ internal degrees of freedom of a site, and (2) it is block-diagonal, consisting of two Hermitian $6 \times 6$ matrices per site.
The Hermitian matrices are stored in a packed format, taking $2\times(6 \text{ real diagonal elements } + 15 \text{ complex off-diagonal elements}) = 72$ real numbers per site.

The major part of computation time in \lqcd simulations is spent in solving linear systems $Au = f$, where $A$ is for example the Wilson-Clover operator.
In the next section we explain how this is done by means of iterative solvers, which require repeated applications of the operator $A$ to a vector.
Applying $A$ involves the following.
In the stencil part $D_{\text{w}}$ the computation of the hopping term for a given direction starts by projecting the 24-component spinor on a site down to a 12-component half-spinor.
The upper and lower halves of the half-spinor are then multiplied by the SU(3) gauge link matrices, which are represented as $3\times 3$ matrices.
Finally, the 24-component spinor is reconstructed from the 12-component half-spinor, and a sum over all 8 directions is performed.
In total, $D_{\text{w}}$ requires $1344$ flop/site.
The Clover term $D_{\text{cl}}$ (and the diagonal term $N_d+m$) multiplies the upper and lower halves of the 24-component spinor by Hermitian $6\times 6$ matrices.
This requires $504$ flop/site.
In total, the operator $A$ performs $1344+504=1848$ flop/site.

\subsection{Iterative solvers and preconditioners}
\label{sec:solvers}

The solution of the large and sparse system $Au=f$ is typically found by an iterative solver such as CG~\cite{Hestenes:1952}, GMRES~\cite{Saad:1986}, or BiCGstab~\cite{vanderVorst:1992}.
These solvers find an approximation to $u$ in the so-called Krylov space, spanned by the vectors $f, Af, A^2f,\dotsc,A^kf$.
This space is constructed iteratively by repeated application of the operator $A$ to a vector. Each iteration also involves an orthogonalization process.
For lattice volumes of interest, limited memory size and the required number of $1848$ flop/site/iteration often force a multi-node implementation of the solver to facilitate convergence within reasonable wall-clock time.
As a consequence, iterative solvers may suffer from two limiting factors.
First, applying the operator $D_{\text{w}}$ requires large amounts of communication with nearest neighbors, so the algorithm can become network-bandwidth (and latency) bound.
Second, the orthogonalization process requires global sums, which can introduce large latencies when the algorithm is scaled to many compute nodes.

A common technique to alleviate these issues is \emph{preconditioning}.
The original system is rewritten as $AMM^{-1}u=AMv=f$, where one should find an $M$ such that $M \approx A^{-1}$.
This new system is then solved for $v$, and the solution to the original problem is given by $u=Mv$.
Since $M$ is chosen such that $AM \approx 1$, the resulting preconditioned matrix $AM$ will be more diagonally dominant than $A$, and hence the iterative inversion with $AM$ will converge faster than the original one with $A$.
In each iteration of the iterative solver for the preconditioned system $AMv=f$, both the preconditioner $M$ and the operator $A$ are applied to a vector.
The remainder of the solver is largely unchanged and still involves orthogonalization of the basis vectors.

A general advantage of preconditioning lies in the option to apply $M$ only approximately, either as an iterative process or in lower precision.\footnote{For example, in single- or half-precision for a double-precision solver, often without significant negative influence on the solver iteration count.}
The iterative solver must be able to accommodate such an approximation to $M$, which will in general not be constant over the whole iteration.
Such solvers are referred to as `flexible'.
Since the preconditioner itself can also involve iterative processes, the solver for $AM$ is usually referred to as \emph{outer  solver} or \emph{outer iteration}.
In our implementation of a solver for the Wilson-Clover operator we use flexible GMRES with deflated restarts~\cite{Frommer:2012zm} as outer solver and the multiplicative Schwarz method~\cite{Schwarz:1870} adapted for the case of QCD~\cite{Luscher:2003qa} as a preconditioner.
The latter is a so-called domain-decomposition (DD) method, which is outlined and motivated in the following subsection.
As we shall see, it leads to a largely reduced amount of communication, without any global sums, and to a higher degree of cache reuse.


\subsection{Domain decomposition}
\label{sec:dd}

\begin{figure}
    \centering
    \includegraphics[width=\linewidth]{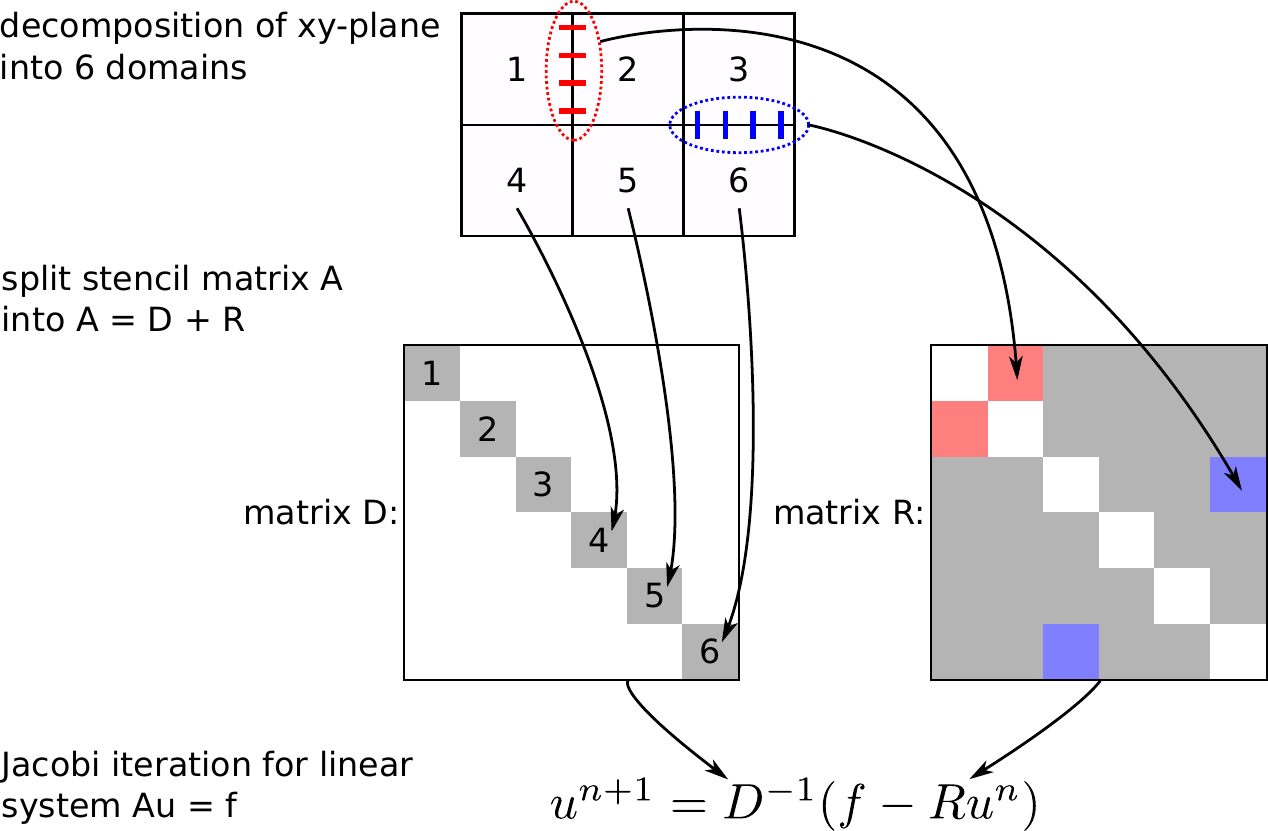}
    \caption{\label{fig:dd}Illustration of Schwarz domain decomposition.}
\end{figure}

We outline the Schwarz method for the system $Au=f$, which is basically a block-Jacobi iteration based on a domain decomposition of the space-time volume.
Our code implements the \emph{multiplicative} Schwarz method, which is an extension of the \emph{additive} method, but for simplicity and brevity we restrict our explanations to the latter variant.
The algorithm proceeds as follows.
The space-time volume is split into a number of domains, and the linear system is reordered such that the degrees of freedom of a given domain have contiguous indices --- this gives the linear system a block-structure.
As illustrated in Fig.~\ref{fig:dd}, the iteration is then based on splitting the matrix $A$ into a block-diagonal part $D$ and the remainder $R$, with $A=D+R$.
The blocks in $D$ correspond to hopping terms within a domain, while blocks in $R$ correspond to couplings between domains.
For a nearest-neighbor stencil, many of the blocks in $R$ are zero.
Based on the splitting into $D$ and $R$ we can write down the (block-)Jacobi iteration for this system,
\begin{align}
    u^{n+1} &= D^{-1}(f-Ru^n)
\end{align}
with $u^0 = 0$.
For certain linear systems this iteration converges to the exact solution $u$, but even in cases where it does not, the approximate solution $u^k$ obtained after a few iterations is often good enough to be used in a preconditioner.
Typically $I_{\mathrm{Schwarz}}=\mathcal{O}(10)$ iterations are performed when the Schwarz method is used as a preconditioner.
In that case the vector $f$ is an input from the outer solver (the current iteration vector) and the preconditioner returns an approximate solution $u^{I_{\mathrm{Schwarz}}} = Mf$ to the outer solver.

The crucial property of the Schwarz iteration is that the computation of $D^{-1}$ can be done \emph{locally} and thus without network communication, since $D$ couples only degrees of freedom within each domain.
The action of $D^{-1}$ on a vector is also computed with an iterative solver. Here, typically $I_{\mathrm{domain}} \lesssim 10$ iterations are sufficient.
As a second benefit, the domain size can be adjusted such that the computation of $D^{-1}$ can be done from cache, without memory access. This can yield a very high performance of this algorithm part, by reducing the dependence on the main-memory bandwidth.

The application of $R$ does require communication, but $R$ is needed only once per inversion of the blocks on the diagonal.
Compared to a direct inversion without preconditioner, the communication is thus roughly reduced by a factor given by the number of block iterations $I_{\mathrm{domain}}$.
As a consequence the solver algorithm with Schwarz preconditioner is better suited for strong-scaling than an inversion without preconditioner.

Finally, the Schwarz method does not use any global sums. There are still global sums for the orthogonalization done by the outer solver, but the frequency of these is reduced greatly, by a factor of $I_{\mathrm{Schwarz}}I_{\mathrm{domain}}$.

We use the minimal residual (MR) algorithm~\cite{Saad:2003} to invert the block-diagonal matrix $D$.
It requires only three vectors for its iteration and thus makes it easier to run the block solve from cache without memory access.
For the inversion on a domain corresponding to a block of $D$ we use even-odd preconditioning, which solves the Schur-complement system $\tilde{D}_{ee} \tilde u_e = \tilde f_e$ based on checkerboarding the lattice into even and odd sub-lattices.
Here $\tilde{D}_{ee}$ is the Schur complement obtained from $D$ after checkerboarding,
\begin{align}
    \tilde{D}_{ee} &= D_{ee} - D_{eo} D_{oo}^{-1}D_{oe},
\end{align}
where the subscripts $ee$, $oo$, $eo$, and $oe$ indicate sub-matrices of $D$ mapping between even ($e$) and odd ($o$) sites.
Typically, even-odd preconditioning yields a speedup of about a factor of two by decreasing the number of MR iterations required to reach a certain accuracy~\cite{Luscher:2010ae}.

\section{Implementation details}
\label{sec:implementation}

Our implementation of the solver is written in C++.
Performance-relevant parts are vectorized and use Intel compiler intrinsics.
Threading uses OpenMP and the multi-node implementation is based on MPI.
An overview of the DD algorithm is given in Table~\ref{alg:dd_solver}.
Due to the complexity, most parts could only be represented schematically (for details, see the references given in the algorithm listing).
Recall that the purpose of the preconditioner is to provide a low-precision approximation.
Therefore, it is sufficient to run the Jacobi iteration in single- or half-precision.
Unless stated otherwise, the descriptions in the remainder of this paper will refer to the single-precision case.

In the following, we discuss our efficient \knc implementation of the DD algorithm
--- for a stencil operator in general or the Wilson-Clover operator in particular.
Data layout and cache management for high single-core performance are treated in Secs.~\ref{sec:implementation_simd} and \ref{sec:implementation_cache}.
Threading is covered in Secs.~\ref{sec:implementation_smt} and \ref{sec:implementation_threading}, and the multi-node implementation in Sec.~\ref{sec:implementation_mpi}.

\algsetup{indent=2em}

\renewcommand{\algorithmiccomment}[1]{{\# #1}}
\begin{table}
    \begin{algorithmic}[1]
        \REQUIRE vector $f$
        \STATE\COMMENT{find $\tilde u$ that solves $Au=f$ up to an error of $\varepsilon_{\text{target}}$}
        \STATE\COMMENT{outer solver (flexible GMRES with deflated restarts~\cite{Frommer:2012zm})}
        \WHILE{error $\varepsilon > \varepsilon_{\text{target}}$}
        \STATE\COMMENT{apply preconditioner $M$ (multiplicative Schwarz method~\cite{Luscher:2003qa})}
        \FOR{$s=1$ to $I_{\text{Schwarz}}$}
        \STATE\COMMENT{block solve on each domain (MR~\cite{Saad:2003})}
        \FOR{$n=1$ to $I_{\text{domain}}$}
        \STATE apply Schur complement operator $\tilde{D}_{ee}$
        \STATE BLAS-level1-type linear algebra (\emph{local} dot-products only)
        \ENDFOR
        \STATE communicate boundary data for $R$
        \ENDFOR 
        \STATE apply operator $A$ (communicates boundary data)
        \STATE BLAS-type linear algebra (e.g., Gram-Schmidt orthogonalization, \emph{global} sums for dot-products)
        \ENDWHILE
        \ENSURE approximate solution vector $\tilde u$
    \end{algorithmic}
    \caption{\label{alg:dd_solver}Overview of the DD solver.}
\end{table}

\subsection{Data layout and mapping to SIMD}
\label{sec:implementation_simd}

\begin{figure}
    \includegraphics[width=\linewidth]{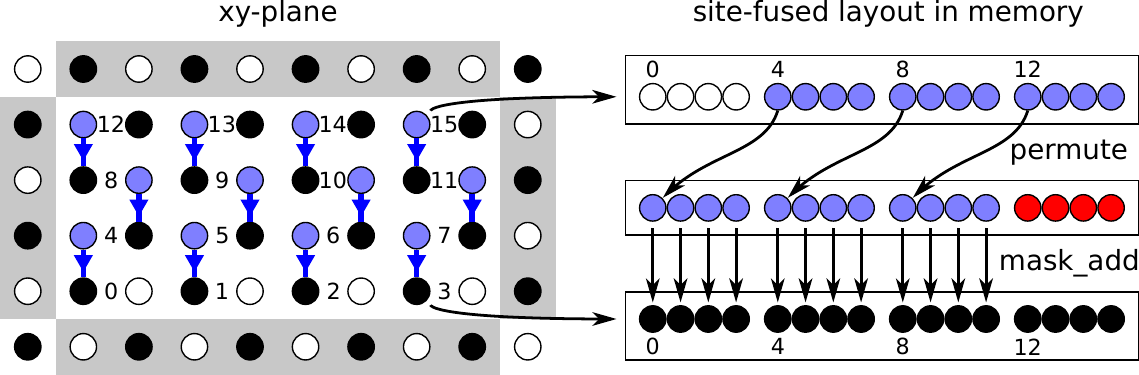}
    \caption{\label{fig:permuting}Illustration of permuting and masking. On the left also parts of the neighbor domains are shown.}
\end{figure}

The data layout and mapping of the algorithm to the wide SIMD units on the \knc is of utmost importance.
In particular we have to (1) avoid loading cache-lines that are only partially needed (such as for terms crossing domain-boundaries), (2) use all SIMD elements wherever possible, and (3) avoid instruction overhead due to permutations.
A detailed consideration of the particular nearest-neighbor stencil of the problem (i.e., the operator $D_{\text{w}}$) shows that the smallest overhead and highest efficiency are obtained when for a given site all components of, e.g., a spinor with its $24$ real degrees of freedom, are stored in $24$ separate registers and cache-lines (known as structure-of-array (SOA) format).
That is, gather and scatter instructions are avoided and there is a 1:1 correspondence between data in memory (cache-lines) and registers.
The elements of the vector registers are thus filled from several sites simultaneously --- $16$ in the case of single-precision. This is known as `site-fusing'.
Furthermore, to facilitate the even-odd preconditioning of the MR iteration, sites from the even and odd checkerboard should not be mixed in the same register.
In combination with cache-size considerations (see Sec.~\ref{sec:implementation_cache}) this lead us to choose site-fusing in $x$ and $y$ directions of $4 \times 4$ sites per even or odd checkerboard.
We refer to these $16$ sites as $xy$-tile.
We illustrate this in Fig.~\ref{fig:permuting} with a cross-sectional view of the $xy$-plane of a domain (white rectangular region on the left).
There is an `even' tile (black circles) and an `odd' tile (white and blue circles).
Numbers are used to label the sites in both the even and odd tile.
The sites of these tiles are interleaved, and together they form the $8 \times 4$ cross section of the domain.

Note that we cannot fuse sites that are far apart, to let the $16$ vector elements handle $16$ independent domains. This would be optimal in terms of computation overhead, but the L2 size is insufficient to hold $16$ domains of reasonable size.

In the $z$ and $t$ directions our vectorization scheme allows for a straightforward computation of all hopping terms
for complete registers of $16$ sites.
Hopping terms in the $x$ and $y$ directions require permutations, e.g., as illustrated for the $y$-forward neighbors (shaded blue) of the even tile (sites shaded black inside the white region) in Fig.~\ref{fig:permuting}.
For the block-inversion the black sites on the $y$-forward boundary do not receive contributions from hopping terms from the $y$-forward direction since they would cross the domain boundary.
To handle this we use the \knc mask feature to disable adding these specific elements (shaded red in the permuted input) of the vector.
This means that for the hopping terms in the $x$ and $y$ directions only $14/16$ and $12/16$, respectively, of the floating-point unit is used, i.e., $12.5\%$ and $25\%$ of the SIMD vectors are wasted.

\begin{figure}
    \includegraphics[width=\linewidth]{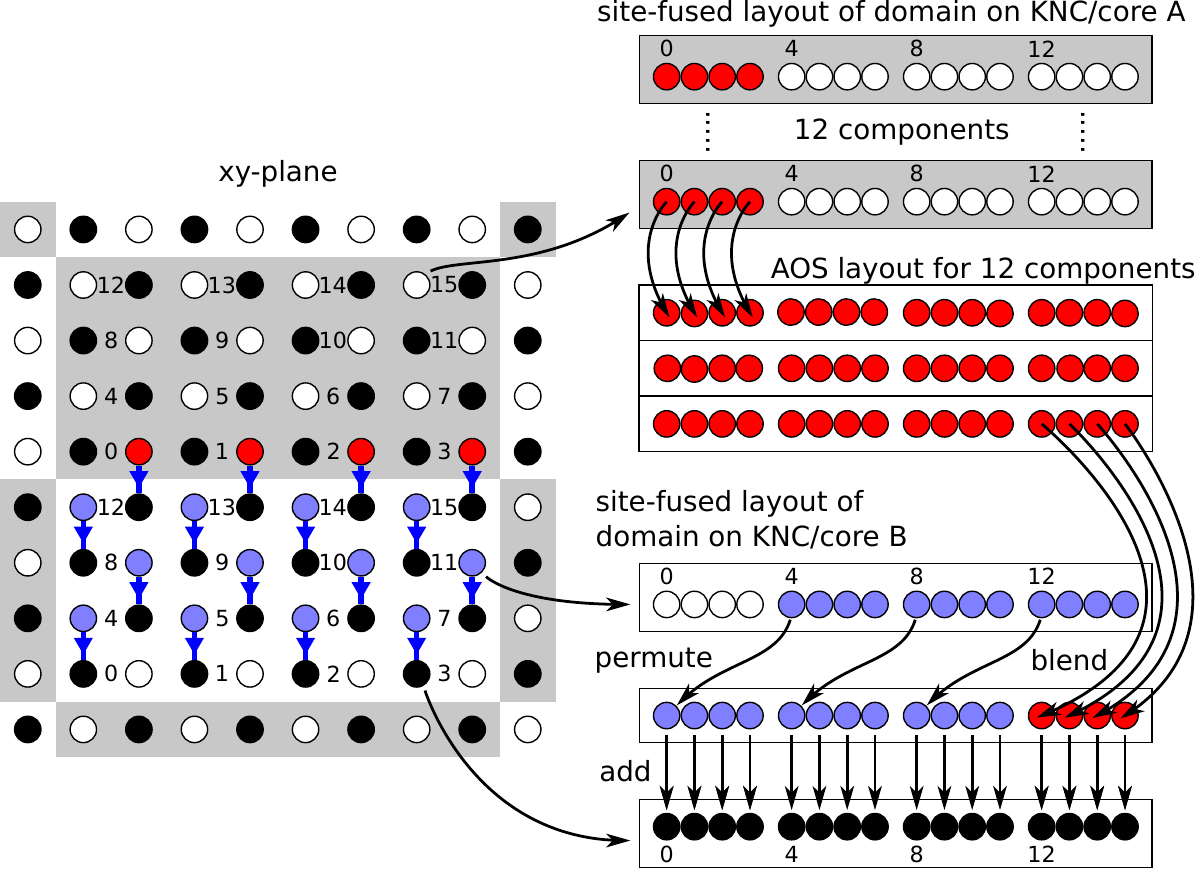}
    \caption{\label{fig:repacking}Illustration of packing boundary data.}
\end{figure}

Before and after the even-odd-preconditioned MR-solve, the full operator $A=D+R$ --- including the hopping terms between domains --- has to be applied to a vector.
If we imagine this from the perspective of a \knc-core working on a given domain, the computation of these hopping terms takes as input the sites lying on the backward or forward boundaries of neighboring domains.
In the $z$ and $t$ directions this works nicely and will access complete cache-lines of fused sites, since each $xy$-tile corresponds to a given $z$ and $t$ index.
In the $x$ and $y$ directions,\footnote{or, in general, in all directions where site-fusing is used} however, only a subset of the sites in a cache-line is on a given $x$ or $y$ boundary --- namely only $2/16$ ($x$ boundaries) or $4/16$ ($y$ boundaries) sites.
This is illustrated on the left of Fig.~\ref{fig:repacking} for the $y$-forward neighbors of the black sites.
There are $12$ internal neighbor sites (marked blue) and $4$ external ones (marked red).
The data for the former are typically available in cache, while those for the latter are not.
Since the \knc always loads complete cache-lines this adds a considerable overhead --- especially since the \knc has no shared last-level cache.
In our implementation a pair of even and odd $xy$-tiles makes up the complete cross section of a domain.
By loading the neighbors that we actually need, i.e., the $2$ or $4$ sites on the boundaries, we would thus end up loading $4$ entire neighbor domains (forward and backward neighbors in $x$ and $y$ directions).

To circumvent this issue we chose to additionally store the boundary data in an array-of-structure (AOS) format before the computation of hopping terms between domains is done.
This is illustrated in Fig.~\ref{fig:repacking}, where the 12-component half-spinors of $4$ sites (red) on the $y$-backward boundary are packed into three cache-lines.
Right after an MR inversion of a domain on a core A, this core packs the relevant half-spinors of sites on the $x$ and $y$ boundaries (needed in the computation of hopping terms) into an AOS format and stores them.
At that point the relevant data are still in cache, so this can be done largely without incurring additional memory traffic.
Furthermore, in many cases it is possible to fuse and/or interleave the packing with the computation of local hopping terms, and extraction for all four $xy$-directions can be done with loading an $xy$-tile only once.
This decreases the overhead added by the packing.
The packed data are later (in the next Schwarz iteration) accessed by a core B, which works on a neighbor domain and inserts the packed data into the $xy$-tiles of that domain.
Thus core B has avoided loading the full data of the neighbor domain.

This is also extremely important for the multi-node implementation, since it avoids an extra iteration over sites (or domains) on the surface for extracting the data to be sent.
Since the extraction is done anyway for inter-domain communication within a \knc, we can simply send these boundary buffers.
In our example, core A would store the data it extracts to a memory location used as a send-buffer.
After exchanging buffer data via MPI, core B --- which can now be on a neighboring MPI-rank --- can read the data from the receive-buffer.


\subsection{Cache management and prefetching}
\label{sec:implementation_cache}

%
The size of the domains is dictated by two factors.
First, the convergence of the method is best with a certain size.
For \lqcd this is typically between $4^4$ and $8^4$~\cite{Luscher:2005rx}.
Since the L2 cache of the \knc is not shared between cores, we restrict each domain to a single core.
Second, for optimal performance, we would like to run the iterative block inversion without exceeding the size of the L2 of a core.
A thorough look at the algorithm (Schwarz method with MR as block inverter with even-odd preconditioning) shows that we need $7$ spinors on the half lattice, in addition to the gauge links and the clover matrices.
Given the L2 size of $512~\mathrm{kByte}$ a domain size of $8\times 4^3$ is practical, since then (in single-precision) the spinors occupy $7\cdot24~\mathrm{kByte}$, the gauge links $144~\mathrm{kByte}$, and the clover matrices $144~\mathrm{kByte}$.
This adds up to $456~\mathrm{kByte}$ and is thus rather close to the L2 size.
To avoid associativity misses it is therefore crucial to pack all required data structures into one contiguous block.

The \knc has no L1 hardware prefetcher, and L2 hardware prefetches can only capture sequential streaming accesses, so software prefetching is crucial.
Due to the domain boundaries the code structure is rather irregular and we cannot rely on compiler-generated prefetches.
Instead, we manually insert L1 and L2 software prefetches into the code (using compiler intrinsics).
Our code is based on many small functions, which execute in several hundreds to a few thousand cycles each.
Consequently, it is inefficient to let each function only prefetch data required by this function, due to cold misses in the beginning of the function.
Therefore, each function also prefetches data for the function that is executed next.

Before the MR inversion of a domain can start, the data associated with this domain must be prefetched from memory into L2.
Ideally this should be done during the MR inversion on the previous domain.
However, the working set of two domains is too large to fit into L2.
Instead, we prefetch the L2 data during the initial phase of the computation on each domain.
We use the code generator from Ref.~\cite{Joo:2013a} to interleave compute instructions with L2 prefetches.
This gives a performance boost with respect to a code version that relies on the hardware L2 prefetcher (see Sec.~\ref{sec:results_single}), but the overall improvement is only moderate --- presumably due to the irregular code structure of the domain-decomposition method.

As an optimization we can furthermore use the \knc hardware support for up- and down-conversion and store domain data in half-precision.
The benefits are a reduced working set as well as reduced bandwidth requirements at all levels of the memory hierarchy.
However, it is not a priori clear how far this would influence the stability of the MR inversion on the domains or the Schwarz method in total, due to overflows or underflows in the conversion.
For the current version of the code we therefore chose to only store gauge links and clover matrices in half-precision, but to keep all spinors in single-precision.
In other words, the vectors in the iteration are kept in higher precision to ensure stability, while the matrix $A$, which is constant during an inversion, is stored in reduced precision.
This reduces the size of both gauge links and clover matrices of a domain from $144~\mathrm{kByte}$ to $72~\mathrm{kByte}$.

\subsection{Intra-core threading}
\label{sec:implementation_smt}

As mentioned in Sec.~\ref{sec:knc}, at least two threads per core are necessary to achieve full utilization of the \knc pipeline.
To avoid pressure on L1 and L2, in our implementation threads on the same core work together on the same domain.
Specifically, threads are assigned to alternating $t$-slices within the domain.
Synchronization and data sharing between threads is fast, since they share the L1 and L2 caches.

In our experience there is no large performance difference between using two or four threads per core.
The former can suffer from more stalls, due to the exposed pipeline latency of L1 or L2 misses. 
However, the latter suffers from more L1 conflict-misses, because our working set exhausts the L1 size.

\subsection{Inter-core parallelization}
\label{sec:implementation_threading}


Since each core works on a domain of its own, threading to use all $N_{\mathrm{core}}$ cores can naturally be done by processing $N_{\mathrm{core}}$ domains in parallel.
A lower bound for the number of sites in the local volume on a \knc is thus $N_{\mathrm{core}}V_{\mathrm{domain}}$, or two times that for the multiplicative Schwarz method where computation alternates between two sets of domains.
If the local volume is smaller, some of the cores will be idle.

In general there are more domains than cores. In that case domains can be assigned to cores in a round-robin manner. Since $N_{\mathrm{core}}=60$, this can lead to load-balance issues for common lattice sizes.
For example, for $256$ domains to be processed, $51$ cores process $5$ domains each, $1$ core processes $1$ domain and idles the rest of the time, and the remaining $8$ cores are idle all the time.
This leads to a load of $256/(5\cdot60) = 0.85$, i.e., $15\%$ of the \knc's compute power is wasted.

There are a few options to reduce the effect of load-imbalance.
The simplest solution is to increase the local volume.
In some cases this may not be an option if time-to-solution is critical (e.g., for Monte-Carlo methods, where a Markov chain is constructed).
In that case a proper choice of the global volume could help.
In particular, prime-factors of $3$ or $5$ in the lattice extent are advantageous, since $N_{\mathrm{core}}=60=5\cdot3\cdot2\cdot2$.
A third approach is a non-uniform partitioning of the lattice, which can make it easier to balance the load.
An example for this is given in Sec.~\ref{sec:results_multi}.

In each Schwarz iteration the computation on a domain, i.e., the MR block solve, is independent of the other domains.
As a consequence each core can work through its set of domains without having to synchronize with other cores.
Before the next Schwarz iteration a barrier among cores ensures that all boundary data have been extracted and stored into the corresponding buffers, and can be used as an input to the next iteration.
The resulting algorithm can thus run for long times without time-consuming synchronization among cores.

\subsection{Multi-node implementation}
\label{sec:implementation_mpi}

A simple way to realize a multi-node implementation is to allow each thread to issue its own MPI calls.
It is easy to implement and avoids the shared buffer management that would be required if MPI calls were issued by only one thread.
However, such an approach has two shortcomings.
First, most MPI implementations incur high overhead when MPI calls are issued in parallel by multiple threads. 
Second, if each core sends individual packets, e.g., for the surface of a domain in a certain direction, these packets are very small, which results in poor bandwidth utilization of the network.


Therefore, our multi-node implementation combines the surface data of all domains and communicates them using a single thread.
Based on a pre-computed offset, cores write the extracted boundary data into a global array for a particular surface.
A \knc-internal barrier ensures that all data have been written, and a dedicated core issues a (non-blocking) MPI-send of this surface to a neighbor rank in this direction.
Similarly, a dedicated core will issue a (non-blocking) MPI-receive and MPI-wait.
A \knc-internal barrier keeps all cores waiting until all data have been received, after which each core can read from the receive buffer with a given offset.
We experimentally confirmed that this approach performs better than MPI calls by individual cores.

As usual, it is important to hide the communication latencies, by overlapping communication with computation.
While the DD method communicates significantly less than a non-DD method, hiding communication can still have considerable impact on performance, especially when the problem is strong-scaled to a large number of nodes.
A standard method for hiding communication is to divide the local volume into surface sites and interior sites.
The computation on interior sites can be done while waiting for data from neighboring MPI ranks, i.e., data which are needed for the computation on the surface sites.
This approach does not work well in our case --- we would need to split the volume into domains on the surface and domains in the interior.
For typical local volumes only few or no domains at all will be in the interior, because domains consist of multiple sites.
For example, for a local volume with $12$ sites in a certain direction and a domain extent of $4$ sites, $2$ domains are on the surface and only $1$ domain in the interior (whereas there are $2$ sites on the surface and $10$ in the interior).


\begin{figure}
    \subfloat[\label{fig:mpi_timesliced}]{\includegraphics[height=0.5\linewidth]{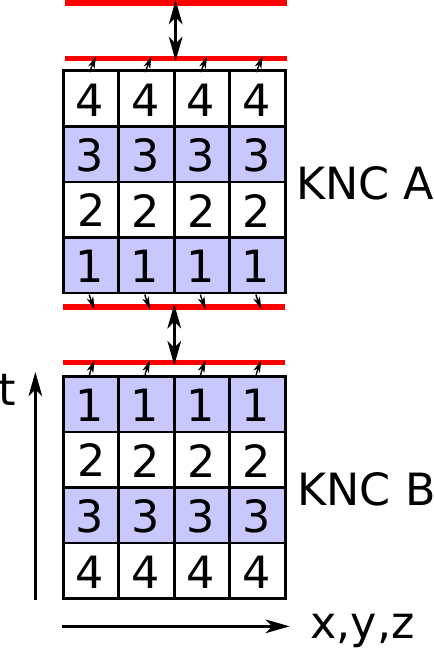}}
    \hfill
    \subfloat[\label{fig:mpi_timesliced_z_split}]{\includegraphics[height=0.5\linewidth]{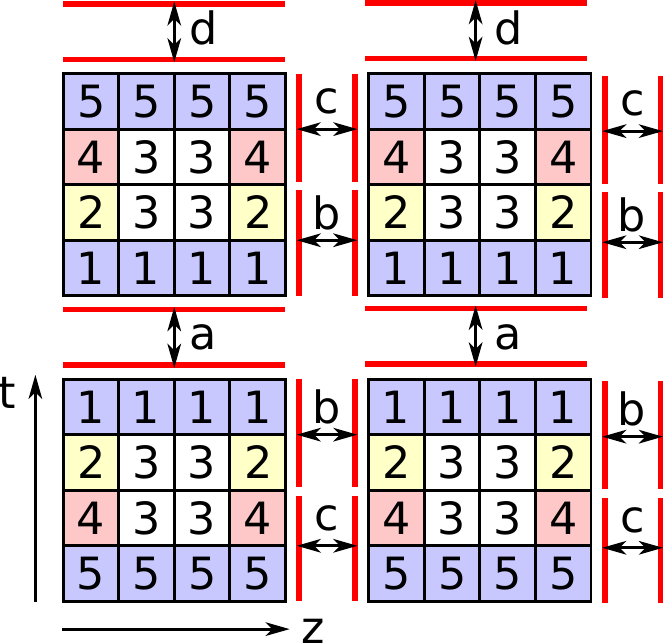}}

    \subfloat[\label{fig:mpi_timesliced_z_split_linear}]{\includegraphics[width=\linewidth]{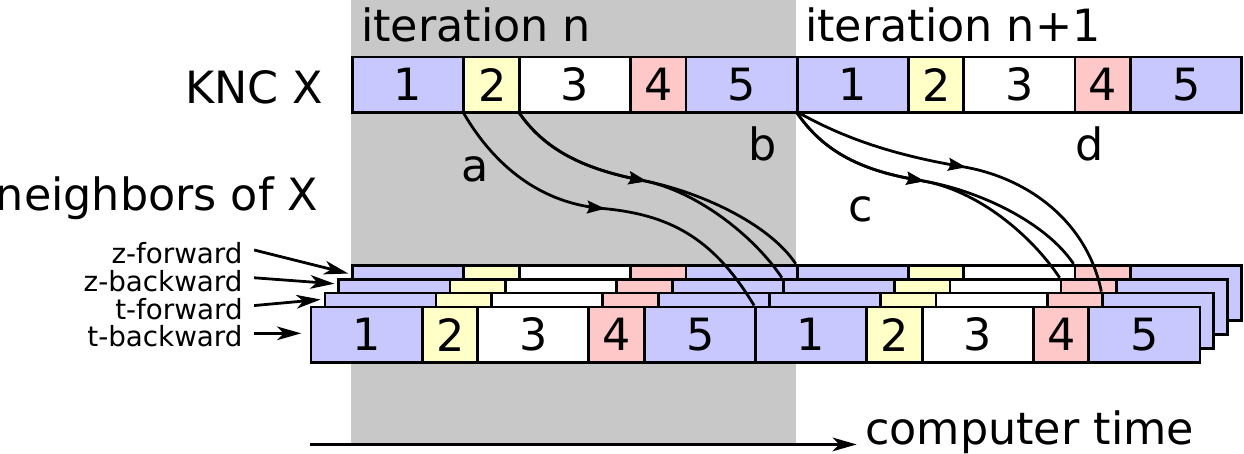}}
    \caption{\label{fig:mpi}Computation pattern for hiding communication. In \ref{fig:mpi_timesliced} and \ref{fig:mpi_timesliced_z_split}, each little square symbolizes a domain. A linear representation of \ref{fig:mpi_timesliced_z_split} is given in \ref{fig:mpi_timesliced_z_split_linear}. It shows the dependencies between computation on various parts.}
\end{figure}
We thus use different methods, as illustrated in Fig.~\ref{fig:mpi}.
In Fig.~\ref{fig:mpi_timesliced}, we show a method which works for splitting in only one direction, here $t$.
Computation proceeds time-slice by time-slice, from (1) to (4). As indicated, a boundary can be sent after computation (1) is done. These data are needed before (1) in the \emph{next} iteration, i.e., communication overlaps with (2-4).
This pattern does not allow to hide communication for $x$, $y$, or $z$ boundaries, because these are ready to send only after (4), but are needed before (1).
Therefore, we devised a new pattern, as shown in Figs.~\ref{fig:mpi_timesliced_z_split} and \ref{fig:mpi_timesliced_z_split_linear}, with splitting in $t$ and other directions.
The computation pattern is as indicated by the numbers. We send $x$, $y$, or $z$ boundaries not after the full boundary data are available, but when \emph{half} of them are ready. As usual, the $t$ boundaries are sent after (1), as indicated by (a). The first half of the, say, $z$-boundary is sent after (2). The second half after (5). The first half will be needed right before the next iteration can start, i.e., the communication (b) is hidden behind computation (3-5). The second half is needed before computation (4) of the next iteration, so (c) is hidden behind computation (1-3) (of the next iteration).

Note that the size of (3), i.e., the interior of the local volume, can actually be zero, but nevertheless our pattern will hide the communication, as long as the number of cores is not larger than half the number of domains.

\section{Results}
\label{sec:results}

\subsection{Hardware and software setup}

Our multi-node benchmarks all use the TACC Stampede cluster.
Each node of the cluster is equipped with a 61-core \knc co-processor (7110P) running at $1.1~\mathrm{GHz}$ with $8~\mathrm{GB}$ memory.\footnote{For our benchmarks we use only 60 cores and stay away from the 61st core where the Linux kernel runs.}
The nodes are connected with FDR Infiniband (Mellanox ConnectX-3 host adapters) with a theoretical peak bandwidth of $7~\mathrm{GB/s}$.
To circumvent hardware issues unrelated to the \knc that would limit the communication performance, we use a previously developed proxy~\cite{Vaidyanathan:2014a} that relays medium and large packets via the host CPU.
The software configuration at the time of our benchmarks was as follows: MPSS version 2.1.6720-21, \knc flash version 2.1.03.0386, Intel MPI version 4.1.0.030, and Intel compiler version 13.1.0.

\subsection{Single-node performance}
\label{sec:results_single}

\subsubsection{Single core}
\label{sec:results_single_core}


As discussed in Sec.~\ref{sec:implementation_cache}, our code can optionally use half-precision gauge links and clover matrices in the preconditioner, while spinors are still stored in single-precision.
Initial tests showed that the use of half-precision has no noticeable negative impact on the overall solver convergence.
For example, in the case of a $48^3 \times 64$ lattice (see below for details) the norm of the residual as a function of iteration count differs by less than 0.14\% between single- and half-precision in the preconditioner.
Therefore, we use this mix of single- and half-precision throughout the benchmarks presented in this paper.

Let us first estimate the upper performance limit.
Since the application of the Wilson-Clover operator $A$ typically takes more than 50\% of the overall time, we only analyze the theoretical bound on its performance. 
As we saw in Sec.~\ref{sec:implementation_cache}, we choose the domain size such that the working set for $A$ fits into the L2 cache.
Therefore, we expect this kernel to be instruction bound rather than memory-bandwidth bound.
In general, the full compute efficiency of the \knc is obtained if the application uses only fused multiply-add instructions.
The Wilson-Clover operator, which is the dominant contribution to the MR iteration, performs 1848 flop/site of which
64\% are fused multiply-adds, giving a maximum achievable efficiency of 82\%.
As described in Sec.~\ref{sec:implementation_simd}, the $x$ and $y$ directions require masking, which causes a 7\% loss
in SIMD efficiency.
Other instructions such as shuffles and permutes also take compute-instruction slots.
Moreover, even though (1) some of the SIMD load instructions can be fused into compute instructions and (2) stores, software prefetches, and most of the scalar instructions for address calculations can be co-issued with compute instructions, not all of them find ideal pairing.
Additionally, stack spills and address computation add to the instruction overhead.
Out of all instructions 54\% are compute instructions.
Of the remaining 46\%, 72\% are pairable and the compiler finds paring for 59\%.
Thus, the compute efficiency is $0.82\cdot0.93\cdot0.54/(1-0.59\cdot0.46)=56\%$.
This gives $(16+16)\cdot0.56 = 18~\text{flop/cycle/core} = 20~\text{Gflop/s/core}$.





For the MR iteration itself we observe a performance of about $12~\mathrm{Gflop/s}$ on a single core.
This is 40\% below the theoretical bound.
The Intel\textsuperscript{\textregistered} VTune\textsuperscript{\texttrademark} performance tool shows that the cause is stalls due to outstanding L1 prefetches (which occur despite aggressive software prefetches).

If we consider the overall time-to-solution, the optimal number of MR iterations is typically small --- for our domain size usually $4$ or $5$.
Other parts in the Schwarz method will thus contribute significantly.
In particular, this involves extraction and insertion of boundary data, computation of hopping terms between domains, and updating the solution and residual vectors.
For the Schwarz method altogether we typically obtain around $8~\mathrm{Gflop/s}$ on a single core.
Table~\ref{tab:single-core} gives an overview of the impact of some of the optimizations discussed in Sec.~\ref{sec:implementation}, namely the performance of the block inversion (MR iteration) and the Schwarz preconditioner (DD method, $I_{\text{Schwarz}}=16$, $I_{\text{domain}}=5$) with and without prefetching and for single- and half-precision gauge links and clover matrices.

\begin{table}
    \centering
    \begin{tabular}{l|cc|cc}
        & \multicolumn{2}{|c|}{MR iteration} & \multicolumn{2}{|c}{DD method} \\
                                   & single & half & single    & half      \\
        \hline\noalign{\smallskip}
        no software prefetching & $5.4$ & $7.9$  & $4.1$ & $5.9$ \\
        L1 prefetches           & $9.2$ & $11.8$ & $5.8$ & $7.7$ \\
        L1+L2 prefetches        & $9.1$ & $11.8$ & $6.3$ & $8.4$ \\
        \hline\noalign{\smallskip}
    \end{tabular}
    \caption{\label{tab:single-core}Single-core performance in Gflop/s. See text for details.}
\end{table}

\subsubsection{Many cores and load balancing}
\label{sec:results_single_knc}

\begin{figure}
    \includegraphics[width=\linewidth]{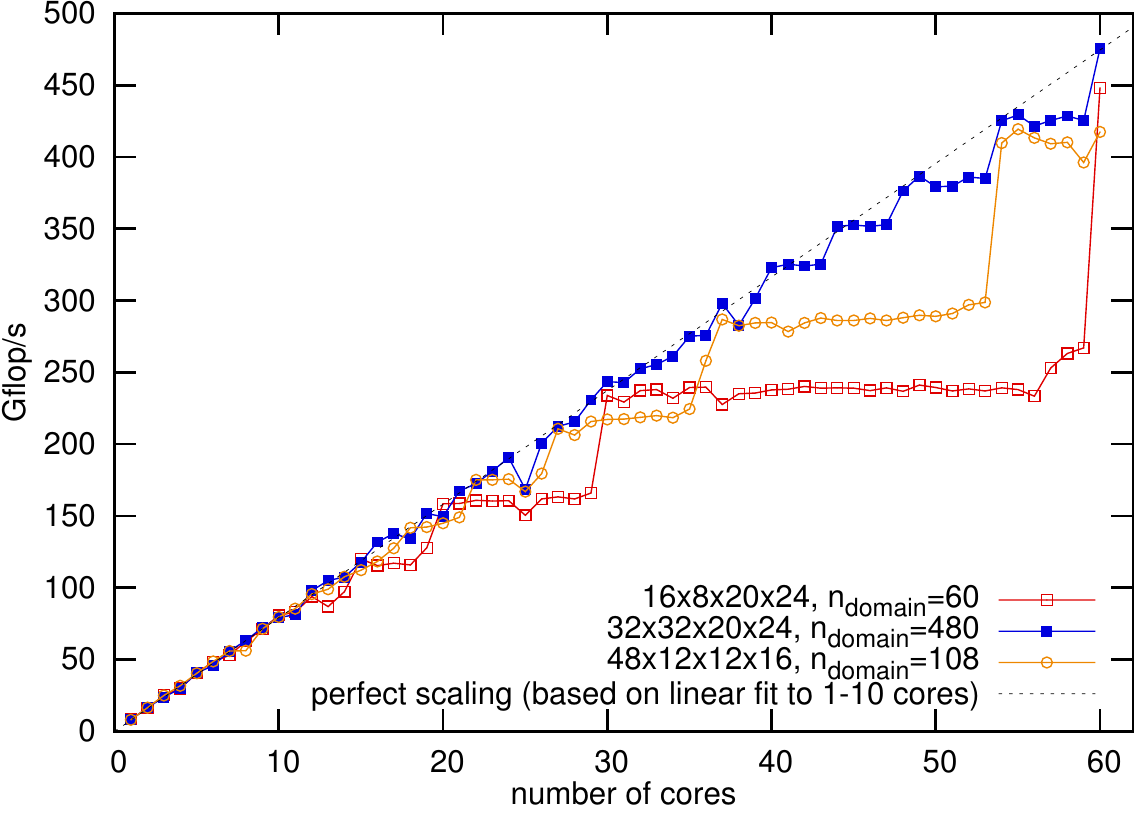}
    \caption{\label{fig:multicore}On-chip strong-scaling of the DD preconditioner ($I_{\text{Schwarz}}=16$, $I_{\text{domain}}=5$). The volumes $16 \times 8 \times 20 \times 24$ and $32 \times 32 \times 20 \times 24$ result in a number of domains that is divisible by 60, so we have $100\%$ load for 60 cores. The $48 \times 12 \times 12 \times 16$ volume corresponds to scaling a $48^3\times 64$ lattice on 64 \kncs, as in Sec.~\ref{sec:results_multi}, with a load average of $90\%$ for 60 cores.}
\end{figure}

As long as a sufficient local volume --- and thus a sufficient number of domains --- is available, the DD method has a very good scaling behavior from 1 core to all 60 cores on a \knc.
In Fig.~\ref{fig:multicore} we show the preconditioner performance as we vary the number of cores.\footnote{These results were obtained with disabled load balancing (set via the cpuset pseudo-filesystem, by setting the flag \verb|cpuset.sched_load_balance| to zero). Load balancing by the Linux kernel causes load on a random core and thus a slowdown of our code by nearly $10\%$. Synchronization among cores then propagates this slowdown to all cores, yielding an overall $10\%$ loss. For these single-\knc tests we were able to control this, but for our multi-\knc tests on Stampede (where load balancing is enabled) we were not.}
Apart from steps due to load imbalance for certain numbers of cores, the preconditioner scales almost linearly with the number of cores.
There are two reasons for this, (1) as discussed in Sec.~\ref{sec:implementation_threading} all cores can work independently, apart from barriers between Schwarz iterations, and (2) 
since a significant part of the DD preconditioner (the MR inversion of the blocks) runs from cache
we do not strongly depend on the memory bandwidth, so even with 60 cores the memory access on the \knc does not become a bottleneck.\footnote{We are using a custom barrier implementation, but since we use barriers scarcely a standard barrier should yield similar performance.}

For a local volume $V$ and domain volume $V_{\text{domain}}=8\cdot 4^3$ the number of domains that can be processed independently (in parallel) is
\begin{align}
    \label{eq:n_domain}
    n_{\text{domain}} &= \tfrac{1}{2} V/(8\cdot4^3).
\end{align}
The factor $1/2$ accounts for the two-color (black/white) checkerboarding in the multiplicative Schwarz method.
The (average) load on $N_{\text{core}}$ cores is then given by
\begin{align}
    \label{eq:load}
    \text{load} &= n_{\text{domain}} / (N_{\text{core}} \cdot \ceil (\tfrac{n_{\text{domain}}}{N_{\text{core}}})).
\end{align}
For the smallest volume in the figure, $16 \times 8 \times 20 \times 24$, and with 60 cores we thus have 100\% load, each core works on one domain, and as a result we achieve linear speedup with 60 cores, as shown in the figure.

\subsection{Multi-node and strong-scaling performance}
\label{sec:results_multi}

\subsubsection{General considerations}
In this section we study the multi-node performance of the DD solver presented in this paper and compare with a solver without DD (labeled `non-DD' hereafter) based on a \knc implementation of the Wilson-Clover operator originally described in Ref.~\cite{Joo:2013a}.\footnote{The work from Ref.~\cite{Joo:2013a} was extended to support the Clover term, to communicate in all 4 dimensions, and to support half-, single-, and double-precision forms of BiCGstab, as well as a mixed-precision solver based on iterative refinement (outer Richardson iteration with BiCGstab as inner solver).}
For a fair comparison of the DD and non-DD solvers, we should consider the wall-clock time for specific cases, such as a solve on a certain number of \kncs.
The wall-clock time is determined by the number of floating-point operations required until the approximation $\tilde u$ to the solution $u$ has reached a certain precision, and the sustained performance in Gflop/s.
Both can differ between solvers and depend on the problem under consideration.
We use the usual definition of the error $\varepsilon$ via the relative norm of the residual, $\varepsilon = \norm{f-A\tilde u}/\norm{f}$.
For all data presented here we use $\varepsilon_{\text{target}}=10^{-10}$, which implies that the (outer) solver uses double-precision.
Additionally, it has to be taken into account that the convergence of a given method strongly depends on the physical parameters.
In the case of QCD this is mainly the pion mass $m_{\pi}$.
We therefore do not compare the performance with fake input data, but with input data from three different production runs (with lattices of size $32^3\times64$, $48^3\times64$, and $64^3\times128$) with parameters commonly used in recent QCD simulations, which are close to nature.\footnote{For reference, we use configurations with 2 degenerate flavors of Wilson-Clover quarks at $\beta=5.29$ ($\beta$ is related to the lattice spacing) and $c_{\mathrm{sw}}=1.9192$~\cite{Bali:2012qs}. For the parameter $\kappa$, which is related to the quark mass, we use $0.13632$ (with a lattice of size $32^3\times64$) and $0.13640$ (lattice size $48^3\times64$) which correspond to pion masses of $m_{\pi}=290~\mathrm{MeV}$ and $150~\mathrm{MeV}$, respectively. The latter is basically at the physical point and will thus make solving of the linear system more difficult than the former.
Furthermore, we use a $64^3 \times 128$ lattice from the USQCD collaboration with 3 degenerate flavors, with a quark mass corresponding to the SU(3) symmetric point~\cite{Durr:2008rw} at $\beta=5.0$ with a non-perturbatively tuned clover coefficient.}


In practice we are interested in two use cases for a solver.
In a first use case, \emph{data generation}, a Markov-chain-based algorithm (typically Hybrid Monte Carlo~\cite{Duane:1987de}) is used, which requires solving many linear systems, one after another.
Building this Markov chain is inherently a serial process, so the strong-scaling limit of the algorithm is of importance for obtaining the longest possible chain within a certain wall-clock time.
In a second use case, \emph{data analysis}, the generated data are analyzed.
This involves the computation of quark propagators, which likewise requires solving linear systems.
The output from different points in the Markov chain can be analyzed independently, i.e., this part of the computation parallelizes trivially.
In this use case, we are thus interested in minimizing the cost for a solve measured in \knc-minutes.

The parameters of the DD method were tuned to yield optimal performance for each lattice.
In particular, for the $32^3\times64$ lattice the outer solver uses a maximum basis size of $8$, with $4$ deflation vectors\footnote{i.e., low-mode vectors kept after GMRES restart,~\cite{Frommer:2012zm}} and we use the Schwarz method with $I_{\text{Schwarz}}=16$ iterations and $I_{\text{domain}}=4$ or $5$ MR iterations.\footnote{With high node count it is sometimes beneficial to increase the MR count in order to decrease the load on the network.}
For the $48^3 \times 64$ lattice we use a maximum basis size of $16$, with $6$ deflation vectors.
Due to the large pion mass of the $64^3 \times 128$ lattice, a maximum basis size of $5$, with $0$ deflation vectors, is sufficient.
In the last two cases we use $I_{\text{Schwarz}}=16$ and $I_{\text{domain}}=5$.

\subsubsection{Strong-scaling (for data generation with Monte Carlo method)}

\begin{figure*}
    \includegraphics[width=\linewidth]{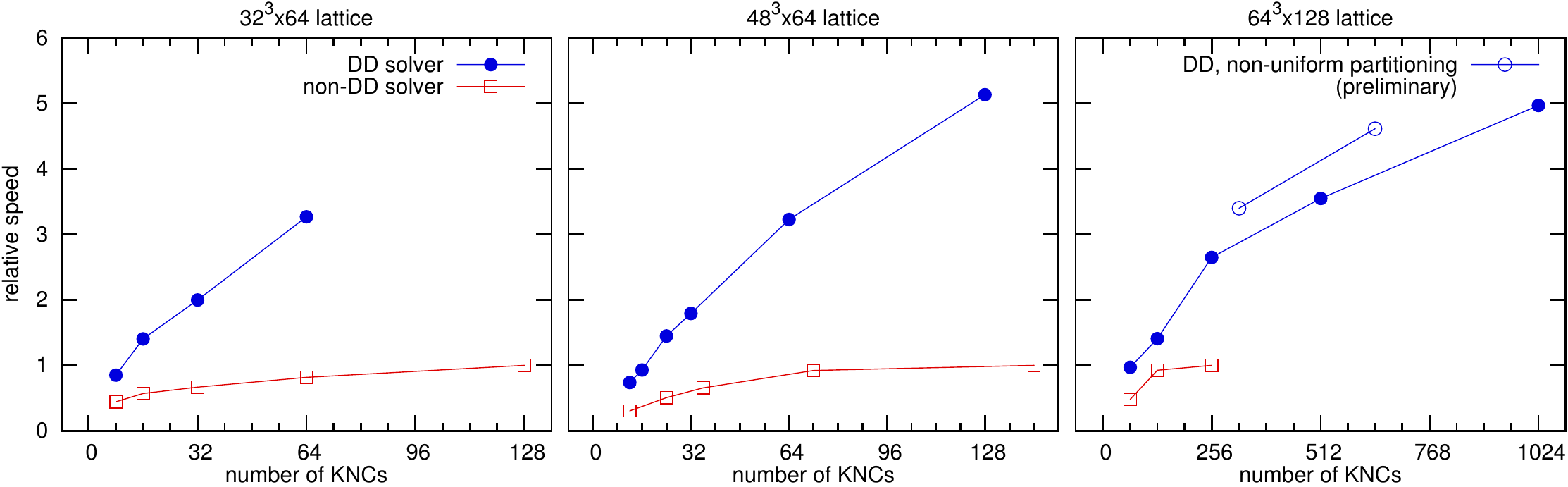}
    \caption{\label{fig:strong-scaling}Strong-scaling: relative speed of the DD and non-DD solvers normalized to the smallest time-to-solution of the non-DD solver.}
\end{figure*}

\begin{table*}
    \centering
    \begin{tabular}{ccc|cccc|cccc|cccccc}
        &&& \multicolumn{4}{|c|}{percent of total time} & \multicolumn{4}{|c|}{Gflop/s/\knc} & & Tflop/s & Tflop/s\\
        \kncs & $n_{\text{domain}}$ & load   & $A$ & $M$  & GS  & other & $A$  & $M$   & GS   & other & iterations & $M$ & total & [s]  & \hspace{-0.6mm}\#global-sums & \hspace{-0.6mm}comm./\knc\\
        \hline\noalign{\smallskip}
        \multicolumn{17}{l}{$48^3\times64$, DD, parameters $m=16$, $k=6$, $I_{\text{Schwarz}}=16$, $I_{\text{domain}}=5$}\\
        24  & 288 & 96\% & 4.3 & 85.8 & 7.8 & 2.1 & 66 & 299 & 56 & 143 & 198 & 7.0  &6.3 & 35.4 & 423 & 15593 \\
        32  & 216 & 90\% & 4.0 & 86.5 & 7.3 & 2.2 & 67 & 276 & 55 & 127 & 198 & 8.6  &7.8 & 28.6 & 423 & 13156 \\
        64  & 108 & 90\% & 4.5 & 85.9 & 6.8 & 2.7 & 52 & 250 & 53 & 92  & 198 & 15.6 &14.0& 15.9 & 423 & 8040  \\
        128 & 54  & 90\% & 5.3 & 83.4 & 7.0 & 4.4 & 35 & 199 & 40 & 42  & 198 & 24.9 &21.6& 10.3 & 423 & 5116  \\
        \hline\noalign{\smallskip}
        \multicolumn{17}{l}{$64^3\times128$, DD, parameters $m=5$, $k=0$, $I_{\text{Schwarz}}=16$, $I_{\text{domain}}=5$}\\
        64              & 512    & 95\% & 4.7 & 89.4 & 3.5 & 2.3 & 64    & 300     & 29    & 24    & 10 & 18.8 & 17.1 & 3.34 & 27 & 488    \\
        128             & 256    & 85\% & 4.4 & 90.0 & 4.0 & 1.5 & 50    & 221     & 19    & 27    & 10 & 27.6 & 25.3 & 2.3  & 27 & 293    \\
        256             & 128    & 71\% & 4.5 & 90.2 & 3.8 & 1.5 & 45    & 204     & 19    & 26    & 10 & 51.0 & 46.8 & 1.22 & 27 & 171    \\
        *320\phantom{*} & 112/64 & 85\% & 4.8 & 89.5 & 4.0 & 1.7 & 48/27 & 230/131 & 20/11 & 26/15 & 10 & 67.3 & 59.9 & 0.95 & 27 & 152/98 \\
        512             & 64     & 53\% & 3.9 & 91.1 & 3.6 & 1.4 & 35    & 135     & 13    & 18    & 10 & 67.5 & 62.7 & 0.91 & 27 & 98     \\
        *640\phantom{*} & 56/32  & 85\% & 4.7 & 88.5 & 5.0 & 1.8 & 33/19 & 158/90  & 11/6  & 17/10 & 10 & 92.4 & 81.4 & 0.70 & 27 & 98/61  \\
        1024            & 32     & 53\% & 5.9 & 86.7 & 4.5 & 2.8 & 16    & 100     & 7     & 6     & 10 & 100.0 & 88.4 & 0.65 & 27 & 61     \\
        \hline\noalign{\smallskip}
        \multicolumn{17}{l}{$48^3\times64$, non-DD: double-precision BiCGstab, SOA-length $=8$ (performs better than mixed-precision solver, since half limited to SOA $=8$)}\\
        12  & - & - & - & - & - & - & \multicolumn{4}{|c|}{entire solver: 70} & 4781 & - & 0.82 & 168.5 & 23,907 & 188,272 \\
        24  & - & - & - & - & - & - & \multicolumn{4}{|c|}{entire solver: 58} & 4777 & - & 1.36 & 101.4 & 23,887 & 115,556 \\
        36  & - & - & - & - & - & - & \multicolumn{4}{|c|}{entire solver: 50} & 4802 & - & 1.77 & 78.4  & 24,012 &  91,848 \\
        72  & - & - & - & - & - & - & \multicolumn{4}{|c|}{entire solver: 35} & 4760 & - & 2.46 & 55.9  & 23,802 &  48,200 \\
        144 & - & - & - & - & - & - & \multicolumn{4}{|c|}{entire solver: 19} & 4728 & - & 2.66 & 51.4  & 23,642 &  26,598 \\
        \hline\noalign{\smallskip}
        \multicolumn{17}{l}{$64^3\times128$, non-DD: mixed-precision Richardson inverter --- outer solver: double (SOA $=8$) --- inner solver BiCGstab: residual $0.1$, single stored as half (SOA $=16$)}\\
        64  & - & - & - & - & - & - & \multicolumn{4}{|c|}{entire solver:           101} & $\approx12\cdot23$ & - & 6.3  & 6.1 & 1408 & 2500 \\
        128 & - & - & - & - & - & - & \multicolumn{4}{|c|}{entire solver: \phantom{1}94} & $\approx12\cdot22$ & - & 11.7 & 3.2 & 1353 & 1314 \\
        256 & - & - & - & - & - & - & \multicolumn{4}{|c|}{entire solver: \phantom{1}56} & $\approx12\cdot24$ & - & 14.1 & 2.9 & 1473 &  948 \\
        \hline\noalign{\smallskip}
    \end{tabular}
    \caption{\label{tab:results}Strong-scaling details. $A$: Wilson-Clover, $M$: Schwarz DD, GS: Gram-Schmidt, other: other linear-algebra in outer solver, comm./\knc: total data sent via the network for full solve (in MB). Lines marked with (*) indicate preliminary results based on a non-uniform partitioning as explained in the text. In that case, one out of five \kncs has fewer domains, and we give numbers for both cases.}
\end{table*}

An overview of the strong-scaling results is given in Fig.~\ref{fig:strong-scaling}, and more details can be found in Table~\ref{tab:results}.
For all tested lattices the DD solver can scale more efficiently to a higher number of nodes and is thus better suited for Hybrid Monte Carlo runs (i.e., the data-generation part of \lqcd).
For example, for the $48^3\times64$ lattice the strong-scaling of the non-DD solver flattens rather early, with a minimal time-to-solution of $51~\mathrm{s}$ with a total of $2.66~\mathrm{Tflop/s}$ reached on 144 \kncs.
The DD solver scales well up to $128$ \kncs, with a minimal time-to-solution of $10~\mathrm{s}$ and a total of $21.6~\mathrm{Tflop/s}$.
At $128$ \kncs there is only a single domain assigned to each core ($6$ cores are unused).
As a consequence, no part of the communication can be hidden.
Nevertheless, the DD preconditioner still runs at nearly $200~\mathrm{Gflop/s}$ per \knc, which is nearly $50\%$ of the single-\knc performance obtained in Sec.~\ref{sec:results_single_knc}.
Given that all of the communication is exposed, this is a very good result,
which primarily due to the fact that the DD solver requires significantly less communication and fewer global sums, as shown in Table~\ref{tab:results}.
In general, the DD solver has a higher single-node performance than the non-DD solver and performs up to $5\times$ faster in the strong-scaling limit.
Furthermore, the DD solver has a higher single-node performance than the non-DD solver.

But let us discuss in detail the performance drop of the DD method when going to more and more nodes.
We have seen in Sec.~\ref{sec:results_single_knc} that a smaller local volume has no large negative influence on the performance --- unless the average load reduces, which is the case for most of the $32^3\times 64$ and the $64^3\times 128$ results, but only marginally so for the $48^3\times 64$ lattice.
Furthermore, Table~\ref{tab:results} shows that the balance between the DD preconditioner $M$ and the other algorithm parts is largely independent of the number of \kncs.
Here, the fraction of time taken by $M$ is always between $80\%$ and $90\%$.
That is, it is not the case that other algorithm parts ($A$ and global sums, which tend to suffer more from strong-scaling) become dominant.
Thus, the main reason for the deterioration of the performance can only be the nearest-neighbor communication.
With increasing number of \kncs it becomes harder to overlap communication with computation, while at the same time the shrinking packet size diminishes the achievable network bandwidth.

We can also observe a slight decrease in performance when going to a larger problem size.
Consider for example the strong-scaling limit of the $48^3\times 64$ and the $64^3\times 128$ lattices with $128$ and $640$ \kncs, respectively.\footnote{Both have similar average load, so a comparison makes sense.}
The problem size (and node count) increases by a factor of $5$, and the performance of the preconditioner $M$ decreases by $20\%$.
Apart from suboptimal tuning of our proxy, this $20\%$ drop may in part be due to the network topology of the Stampede cluster.\footnote{A 2-level Clos fat tree topology. We have no influence on which nodes are assigned to a job, but our results are reproducible over many runs without large fluctuations, so we believe this has no strong impact.}
We plan to test our solver on machines with a torus network, which is better suited for nearest-neighbor communication.
As the number of \kncs increases, the local sub-volume size and hence the number of domains decreases, so that eventually the latter becomes significantly smaller than the number of \knc cores.
This results in a loss of scalability due to decreasing utilization of the \knc resources.
The scalability limit due to underutilization is at $64$, $128$, and $1024$ \kncs for the $32^3\times64$, $48^3\times64$, and $64^3\times128$ lattices, respectively, beyond which the utilization would drop below 50\%.
The scalability could be improved by using smaller domains --- at the expense of increased overhead and most likely a decreased single-\knc performance.
Choosing an optimal domain size is application-specific and is part of our future research.




The low average load (see Table~\ref{tab:results}) for certain volumes ($53\%$ for the $32^3\times 64$ and $64^3\times 128$ lattices in the strong-scaling limit) is mostly an artifact of the uniform distribution of the full lattice to the individual \kncs (done by the QDP++ framework~\cite{Edwards:2004sx} in our implementation).
We can improve upon this with a non-uniform partitioning.
Let us consider the $64^3\times 128$ lattice.
When scaled to $1024$ \kncs our layout of compute nodes is $4\times 4 \times 8 \times 8$, Eq.~\eqref{eq:n_domain} yields $n_{\text{domain}} = 32$, and Eq.~\eqref{eq:load} yields a load of $32/60 = 53\%$.
Here the $t$-direction is split into $8\times 16$ sites.
If we instead split $128$ as $4\times 28 + 16$, i.e., most nodes have $28$ sites in the $t$-direction (resulting in $56$ domains on these nodes), and some have $16$ (resulting in $32$ domains), we can obtain a better overall load.
The average load after this redistribution is $(4\cdot56+32)/(5\cdot60)=85\%$.
This does not increase the overall performance, but significantly decreases the number of \kncs needed for obtaining similar performance --- here $640$ instead of $1024$ \kncs.\footnote{The performance will typically decrease slightly, because the redistribution increases the size of the boundaries that are communicated per \knc.}
We included results of an early test of this approach in Fig.~\ref{fig:strong-scaling} and Table~\ref{tab:results}. We plan to investigate this more thoroughly in the future.
Even more general distributions of the lattice would allow for close to $100\%$ load, but in view of synchronization and communication-buffer management it seems more efficient to restrict the local volume to be a hyper-rectangular subset of the global volume.

\subsubsection{Minimum cost (for data analysis)}

\begin{figure}
    \includegraphics[width=\linewidth]{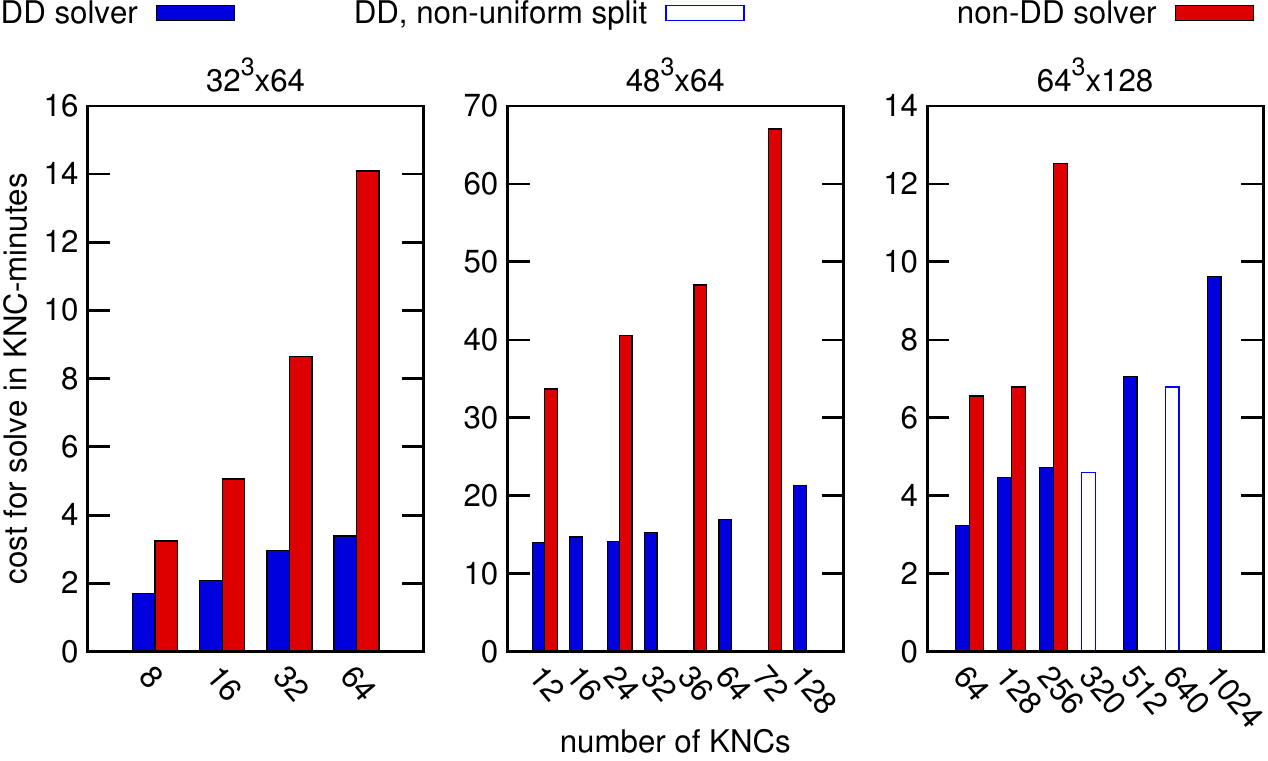}
    \caption{\label{fig:multi-node}\knc-minutes consumed for complete solve.}
\end{figure}

In Fig.~\ref{fig:multi-node} we compare the two solvers based on the cost for a solve, given in \knc-minutes.
As explained above, this is the relevant quantity for data analysis.
Running on as few nodes as the memory footprint allows minimizes the negative impact of network communication and thus yields the lowest cost.
In all cases a solve with the DD method costs about half as much as the non-DD method.

%
%
%



\section{Related work}
\label{sec:related}

DD approaches were first applied to \lqcd by Lüscher~\cite{Luscher:2003qa} using GCR~\cite{Saad:2003} as outer solver, whereas we use flexible GMRES with deflated restarts~\cite{Frommer:2012zm}, which converges faster for problems with low modes.
Other recent work on solvers in \lqcd focused on algebraic multi-grid based and related methods~\cite{Babich:2010qb,Osborn:2010mb,Frommer:2013fsa,Frommer:2013kla,Luscher:2007se}, which can make effective use of DD for their smoothers~\cite{Frommer:2013fsa,Frommer:2013kla,Luscher:2007se}.

Due to their reduced communication requirements, DD methods are particularly suitable for accelerators/co-processors and have been applied with success on GPUs~\cite{Babich:2011:SLQ:2063384.2063478,Osaki:2010vj}.
To the best of our knowledge our present work is the first DD implementation on the \knc and includes several original features relating to this architecture, such as choosing the domain size so that the preconditioner can be applied from L2 cache, thus freeing us from main-memory-bandwidth limitations, and choosing the domain dimensions to facilitate vectorization.
Hence, unlike in Ref.~\cite{Joo:2013a}, we do not explicitly need to worry about tuning the inner length of our structure of arrays, based on the lattice volume.
Unlike GPUs, the \knc does not offer 16-bit arithmetic.
However, we use the precision up- and down-conversion capabilities of the \knc architecture to store the data for the preconditioner in reduced precision.

In an attempt to compare the KNC results reported in this paper to GPU results~\cite{Babich:2011:SLQ:2063384.2063478} we have performed preliminary studies on the NCSA Blue Waters system (Cray XK-7 nodes with Nvidia K20X GPUs).
An important caveat is that exactly identical implementations are not currently available on the two architectures.
Although both implementations use domain decomposition, there are still algorithmic differences (additive vs. multiplicative Schwarz, choice of domain size: fixed vs varying, outer iterative process: GCR vs flexible GMRES with deflated restarts), and hence the comparison cannot be completely rigorous.
We used the same configurations as in the present paper and tried to match the algorithmic parameters to the best of our abilities.
This resulted in quite similar performances.
The run times of the GPU and \knc implementations were within 20\% of each other, with the advantage going one way or the other varying with cases.
Note that the K20X GPU has twice the single-precision peak of the \knc.
The observation that the sustained performance on both architectures is comparable can be attributed to the DD solver's ability to benefit from the much larger caches on the \knc.

For \lqcd the Xeon Phi is clearly superior to (any) dual-socket Xeon in terms of both sustained performance and power-performance (assuming, as is standard in \lqcd, optimal implementations on both architectures).  This was shown for the non-DD solver in Ref.~\cite{Joo:2013a} and is expected to be the case also for the DD solver.










\section{Conclusions}
\label{sec:conclusion}

In this paper, we presented an implementation of an iterative solver with domain-decomposition preconditioner for a stencil operator on \knc clusters.
In particular, we considered the Wilson-Clover operator from the context of \lqcd.
Domain-decomposition methods reduce the amount of data movement from and to main memory as well as via the network, and we could thus make efficient use of the high floating-point performance of the \knc architecture.
For state-of-the-art problem sizes our solver scales to up to 1024 \kncs, with a sustained single-precision performance of around $100~\mathrm{Tflop/s}$ in the preconditioner.
In comparison to a solver implementation without domain decomposition we could reduce the time-to-solution in the strong-scaling limit by a factor of around 5.
The cost for a solve given in \knc-minutes could be reduced by a factor of around 2 when running on few nodes.\footnote{Software and workloads used in performance tests may have been optimized for performance only on Intel microprocessors.  Performance tests, such as SYSmark and MobileMark, are measured using specific computer systems, components, software, operations and functions.  Any change to any of those factors may cause the results to vary.  You should consult other information and performance tests to assist you in fully evaluating your contemplated purchases, including the performance of that product when combined with other products. For more information go to http://www.intel.com/performance.}

A key message of our paper is that to achieve high performance on
modern hardware architectures, a holistic effort is mandatory:
starting from the hardware characteristics, we selected the most
appropriate algorithm.  We adapted the data layout to make optimal use
of the hardware (cache and vector units) and to minimize communication
to external memory and the network.  We carefully tuned the most
relevant kernel(s), including cache management and prefetching,
intra-core threading, inter-core parallelization, and multi-node
implementation.  We also employed further optimizations, such as the
use of mixed precision (if beneficial) and tricks to hide
communication latencies.  Many of the techniques discussed here should
also be applicable in other computational methods that use stencil
operators.

Our current implementation still leaves some room for various further code optimizations, but apart from that there are several options for a potential further improvement of the solver performance.
Smaller domains could be used to push the strong-scaling limit further, by decreasing the minimum per-core volume.
Provided that there are no stability issues, we would like to exploit using half-precision also for the spinors in the domain-decomposition preconditioner.
This would reduce the requirements on cache size, memory bandwidth, and network bandwidth.
Also, the outer solver could be implemented in mixed-precision (single- and double-precision), to further reduce its (already small) contribution to the total solve time.
This would allow us to do most of the linear algebra for basis orthogonalization and the operator application in single-precision.
We will explore these possibilities in the future.

Intel recently announced details of the next generation of {Intel\textsuperscript{\textregistered} Xeon Phi\textsuperscript{\texttrademark} products, which
is called `Knights Landing' (\knl).  Its floating-point performance and
memory bandwidth will be about three times higher than on the \knc so
that the sustained performance (in \% of peak) is expected to be
roughly the same.  The improvements of the compute cores (such as
improved branch prediction and L1 hardware prefetching) will make
low-level optimizations easier to program.  Porting our code from \knc
to \knl will require only modest efforts since the instruction set
architecture is quite similar, and hence the maintainability of the
code base is ensured.

\section*{Acknowledgements}

This work has been supported by the Deutsche Forschungsgemeinschaft (SFB/TR 55).
The authors acknowledge the Texas Advanced Computing Center (TACC) at The University of Texas at Austin for providing HPC resources that have contributed to the research results reported within this paper. 
B. Joó acknowledges support through the SciDAC program funded by the U.S. Department of Energy, Office of Science, ASCR, NP, and HEP Offices and through U.S. DOE Contract No. DE-AC05-06OR23177 under which Jefferson Science Associates, LLC operates Jefferson Lab.
The U.S. Government retains a non-exclusive, paid-up, irrevocable,  world-wide license to publish or reproduce this manuscript for U.S. Government purposes.
The $64^3\times 128$ lattice field configuration was produced  on the Blue Waters computing system, supported by the U.S. National Science Foundation (award number ACI 1238993) and the state of Illinois. 
The configuration was generated  as part of research under PRAC project `Lattice QCD On Blue Waters' award no. OCI08-32315.



\bibliographystyle{IEEEtran}
\bibliography{dd_knc}
%
%
%

\end{document}